\documentclass[aps,prl,twocolumn,superscriptaddress,showpacs]{revtex4}

\usepackage{graphicx}

\begin{document}
\title{Spreading and localization of information in scale-free networks with communication constraints}
\author{Kosmas Kosmidis} \author{Armin Bunde}
\affiliation{Institut f\"ur Theoretische Physik III, Justus-Liebig-Universit\"at
  Giessen, 35392 Giessen, Germany}
\date{\today}
\begin{abstract}
We study localization of information on scale free networks with communication constraints when, for some reason, information can propagate only between ``mutually trusted nodes'' (MTN). We propose an algorithm to construct the MTN network and show that there is a critical value $\bar{m}_{c}>2$ of trusted nodes below which information localizes.  This critical value increases drastically if a fraction $p$ of nodes does not transfer information at all. 
We study the fraction of initial messangers needed to inform a desired fraction of the network as a function of the average number of trusted nodes $\bar{m}$ and discuss possible applications of the model to marketing, to the spreading of risky information and to the spreading of a disease with very short incubation time.

\end{abstract}
\maketitle

In scale-free networks, the number of links (``degree'') emanating from a node follows a power law distribution, $ P(k)\sim k^{-\gamma} $, and the observation that a large number of natural networks, in particular social and biological networks, follow this distribution (with $2<\gamma<3$) \cite{AJB,BAL,Amaral,BALRev,Mendes,NWS} has stimulated research tremendously. Scale-free networks exhibit the ``small world'' phenomenon \cite{Coh1}, an effect which had been discovered, already 4 decades ago, by Milgram when studying the separation between individuals in the USA (``six degrees of separation'')\cite{Milg,w1}. The broad degree distribution of scale-free networks implies the presence of a capable number of well connected nodes (hubs) that decisively influence the properties of the network.
Due to the hubs the network is robust against random distraction of its nodes \cite{Cohen}, favoring e.g. the spreading of information or epidemics \cite{Newm,Gallos,Vesp}, even if the network is widely immunized. The percolation and epidemics threshold on a scale-free network (with $2<\gamma<3$) is zero even in the case of clustered networks \cite{SerBog}.

As the route from one person to another is ultra-small, we expect information and epidemics to spread very fast. A well known example where information flowed as expected is the Yom-Kippur war, where news spreaded with very high speed without radio or TV transmissions. But there are other cases where highly relevant information might have remained localized contrary to any intuition. There are still conflicting opinions on what the vast majority of people in Europe, and in particular in Germany, knew \cite{Dahn} about the concentration camps during World War II  and  it is not clear how well the people in the former Soviet Union was informed about the gulags. In Marketing, there are examples of products that become famous with minimal advertisement while others remain unknown, and only epidemics with a sufficiently large incubation time (like AIDS) seem to spread as expected.  Other cases of practical interest may include information of economic nature, for example about a future currency devaluation or a merger between two companies. Motivated by these examples, we study here the condition for information localization to occur.

We examine spreading phenomena in a scale-free network when the transfer between nodes is constrained. Communication constraints in a network exist when it is not beneficial for a node to share information with all its neighbors, either because the information is risky or because the messanger has limited resources or time to spread the information and thus, has to select some of his neighbors and transfer information only to them. As a pedagogical example, we focus on the spreading and localization of information that is relevant to everybody and risky for the messanger. We do not deal with ``specific signaling'' \cite{Sneppen} as it is not relevant to everybody or rumor spreading \cite{Moreno,Zanette} as, in this case, the concept of risk is absent.

We assume that when someone has information which is dangerous for him to disclose, he will not reveal it to all of his acquaintances but only (on the average) to those $\bar{m}$ nodes of his neighbors that he trusts at most. We find that there exists a critical value $\bar{m}_{c}$ of ``trusted nodes'' which separates information spreading from information localization. Above $\bar{m}_{c}$ information spreads while below $\bar{m}_{c}$ it is localized. This information percolation threshold is quite higher than intuitively expected and can increase drastically when in addition a certain fraction $p$ of nodes is not willing to transfer information at all.

\begin{figure}
\begin{center}
\includegraphics[height=8.5 cm,width=8.5cm]{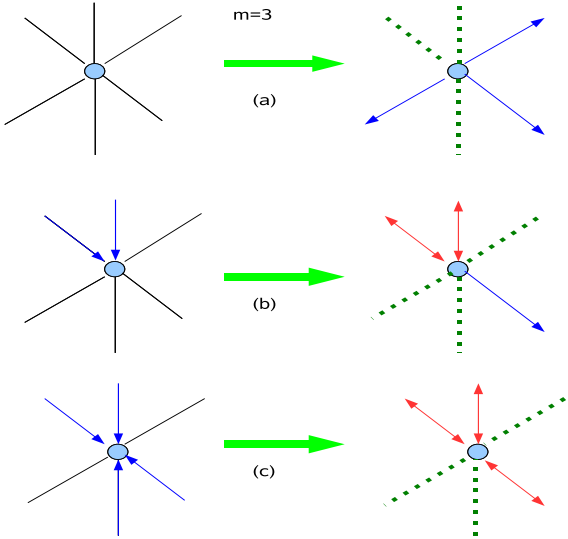}
\end{center}
\vspace{-1.5cm}
\caption{Schematic illustration of the creation  of the ``mutually trusted node'' (MTN) network, when the number of trusted nodes $m=3$. For convenience, only the ``correlated ''( $q=1$) case is shown. 
We consider a certain node $i$ with 6 links (black lines). In (a) the node has not been chosen by another node as trusted , i.e there are no friendly nodes. In (b) there are two (less than $m$) friendly nodes characterized by the blue single arrows, while in (c) there are four (greater than $m$) friendly nodes.
In (a) three of the neighbors are chosen as trusted (single blue arrows) but mutually trusted bonds are not yet established.
In (b) the two friendly nodes are chosen with probability $q=1$ establishing mutually trusted bonds (red double arrows) and a third trusted node is chosen (single blue arrow). Other links are rendered inactive (green dotted lines).
In (c), three of the four friendly neighbors are chosen as trusted nodes, establishing three mutually trusted bonds (red double arrows). The dotted links remain inactive forever.
Information propagates only through the network of ``mutually'' selected nodes, the MTN network (red double arrows). }
\label{fig1}
\end{figure}

In a network, individuals are presented as nodes and if two individuals have established a relationship then they are linked, otherwise the link is absent.  It has been shown recently, that scale-free networks with $\gamma \sim 2.5 $ yield a reasonable description of the science collaboration or actors networks \cite{Amaral}, and a model of mobile agents has been proposed for the formation of this type of social networks \cite{Herrm}. In our analysis, we consider scale-free networks with three values of $\gamma$, $\gamma = 2, 2.5$ and 3 and minimum degrees $k_{min}=1$ and 2. For comparison, we also consider a real social network of an internet dating community (IDC network) \cite{HEL}.
Our basic assumption is that risky information can propagate only between ``mutually trusted nodes'', since risky information transfer needs (i) trust of the receiver in the messanger for believing in it and (ii) trust of the messanger in the receiver for not disclosing the information to unwanted individuals.

\begin{figure}
\begin{center}
\includegraphics[width=8.5 cm,angle=0]{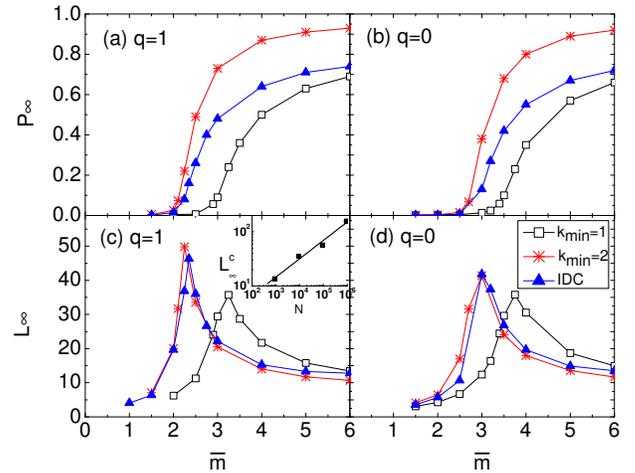}
\end{center}
\vspace{-0.8cm}
\caption{ Information localization transition: (a) Fraction of the largest cluster $P_{\infty}$   vs $\bar{m} $ (average number of trusted nodes), for a random scale free network with $N=10^{4}$, $ \gamma=2.5 $, $k_{min}=1$ (open squares) and $k_{min}=2$ (stars),respectively, and an internet dating community network (IDC)with $ N=14782 $(filled triangles) \cite{HEL}  for $q=1$. The critical values are: $\bar{m}_{c}= 2.1$ ($\gamma=2.5$ , $k_{min}=2$), $\bar{m}_{c}=2.3$ (IDC network) and $\bar{m}_{c}=2.8$ ($\gamma=2.5$, $k_{min}=1$) (b) Same as (a) but for $q=0$. In this case, $\bar{m}_{c}= 2.7$ ($\gamma=2.5$ , $k_{min}=2$), $\bar{m}_{c}= 2.9$ (IDC network) and $\bar{m}_{c}=3.2$ ($\gamma=2.5$, $k_{min}=1$) . (c) Average distance ($ L_{\infty}$) of the nodes on the largest cluster vs $\bar{m}$ for $q=1$ for the same networks as above (d) Same as (c) but for $q=0$. \emph{Inset}: Log -Log plot of $L_{\infty}^{c}$ vs the network size. The slope of the straight line is 1/3. }
\label{fig2}
\end{figure}

By definition, the network of mutually trusted nodes (MTN-network) is a subnetwork of the original network. For constructing the MTN network sequentially, we consider either a given social network or an artificial scale-free network of $N$ nodes with a certain exponent $\gamma$ and a certain minimum degree $k_{min}$. In each step, we choose randomly a certain node $i$ with $k_{i}$ links that has been selected already by $s_{i}\leq k_{i}$ neighboring nodes (``friendly neighbors'') as trusted, and choose, from a Gaussian distribution of mean $\bar{m}$ and width $\sigma$ its number of trusted neighbors $m_{i}$. If $k_{i}\leq m_{i}$, all neighboring nodes are selected. If $k_{i}>m_{i}$, there are different ways of selection that depend on the weight the friendly neighbors are given (see Fig. \ref{fig1}). In the case of ``uncorrelated choice'', node $i$ chooses its trusted neighbors randomly, each one with probability $p_{i}=m_{i}/k_{i}$. In the case of ``correlated choice'', node $i$ chooses its trusted neighbors  in a correlated way such that the friendly neighbors are selected with a higher probability. First, only the $s_{i}$ friendly neighbors are considered, and selected with probability $q$ until the desired number of $m_{i}$ trusted nodes is reached. If this is not the case yet, then, the remaining number of trusted nodes is chosen randomly from all neighbors $k_{i}$ that have not yet been selected. 
Figure \ref{fig1} shows an illustration of the selection process for the fully correlated case $q=1$. A characteristic property of the model is that mutual trust between two hubs is difficult. This can easily be seen in the ``uncorrelated'' case $q=0$, where the probability that a node $i$ with degree $k_{i}$ and a node $j$ with degree $k_{j}$ both select each other as a trusted neighbor is proportional to $\frac{1 }{k_{i} k_{j}}$ and becomes negligible if the two nodes are hubs.
\\ The mean number of trusted neighbors $\bar{m}$ per node is the characteristic tunable parameter in the system. We intuitively expect that for low $\bar{m}$ values the MTN network will be fragmented and information will be localized while for high $\bar{m}$ values (above a critical value $\bar{m}_{c}$) a giant component will appear where information can spread. 
In the following, for the sake of simplicity, we will focus on the uncorrelated  model ($q=0$) and the fully correlated model ($q=1$). We are interested in the transition between localization and spreading of information.

Figure \ref{fig2} shows the fraction of nodes $ P_{\infty} $
\footnote{It can be shown that, within the mean field theory $P_{\infty}$ is the square root of the ``connectivity measure'' $C$ of a social network, used in sociology to describe the fragmentation of a society (S. Havlin, private communication).}
that belong to the largest cluster as a function of $\bar{m}$ for both artificial and real networks for (a) $q=1$ and (b) $q=0$. The size of the largest cluster is, by definition, the maximum number of nodes that eventually learn a risky information from one randomly selected node. This fraction is negligible for $\bar{m}$ below a critical threshold $\bar{m}_{c}$ indicating information localization, while there is a sharp increase of $P_{\infty}$ above $\bar{m}_{c}$ indicating information spreading. The figure remains qualitatively the same for networks of different $\gamma$ values in the range $2<\gamma<3$. The position of the threshold $\bar{m}_{c}$ can be more accurately determined through the condition $ \left\langle  k^{2} \right\rangle  / \left\langle k \right\rangle   =2$, where $k$ denotes the degree of a vertex and $< >$ stands for the mean value over all graph nodes \cite{Cohen}. 
For a scale-free network with $\gamma=2.5 $ and $k_{min}=1$, we find $ \bar{m}_{c} =2.8$ for $q=1$ and $ \bar{m}_{c} = 3.2$ for $q=0$. For $\gamma=2.5 $ and $k_{min}=2$, we find $ \bar{m}_{c} =2.1$ for $q=1$ and $ \bar{m}_{c} = 2.7$ for $q=0$.
For the IDC network, $\bar{m}_{c}$=2.3 for $q=1$ and 2.9 for $q=0$  
(see also \footnote{We applied our model to four small records of criminal acquaintance networks. Confidentiality is  important in such  networks, which are quite different from an internet dating community network. Thus, we expect different results. Indeed, for half of these networks, $\bar{m}_{c}$ was above 6 while for the other half it was not possible to have percolation of information at all.}). 

\begin{figure}
\begin{center}
\includegraphics[width=8.5cm]{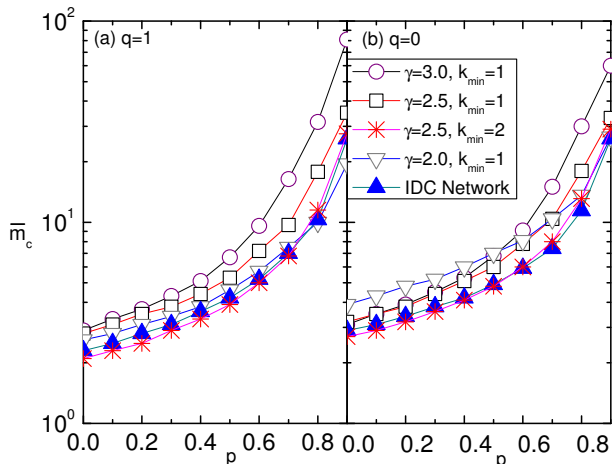}
\end{center}
\vspace{-0.8cm}
\caption{Phase diagrams for (a) $q=1$  and (b) $q=0$. The log - linear plot shows the critical value of $\bar{m}$  vs the probability $p$ that a node does not share the information with its trusted neighbors. The curves separate a phase of ``information localization'' (below the curve) from a phase of ``information spreading'' (above the curve). Curves are for four random scale free networks with $N=10^{4}$ nodes and $\gamma=2 $, $ k_{min}=1$ (open triangles); $\gamma=3$, $ k_{min}=1$ (open circles); $\gamma=2.5$, $ k_{min}=1 $ (open squares); $\gamma=2.5$, $ k_{min}=2 $ (stars); and for the internet dating community network (IDC) with $ N=14782 $ (filled triangles).} 
\label{fig3}
\end{figure}

While $P_{\infty}$ is the maximum fraction of nodes that can be informed by one node, the mean time $t_{\infty}$ to reach all possible nodes is proportional to the average topological distance $L_{\infty}$ between them. Figure \ref{fig2}(c) and (d) show  $L_{\infty}$, as a function of $\bar{m}$, for the same networks as in Fig. \ref{fig2}(a) and \ref{fig2}(b), again for the correlated ($q=1$) and the uncorrelated ($q=0$) model.
For low values of $\bar{m}$, where the network is fragmented, $P_{\infty}$ is small and so is $L_{\infty}$. 
For large values of $\bar{m}$, where the network is well connected, $L_{\infty}$ is again small, exhibiting the “small world” effect. Near $\bar{m}_{c}$ we observe a sharp increase, indicating the complex structure of the resulting network.
In the inset we examine how $L_{\infty}$ scales with the system size. We find a power law scaling with exponent close to 1/3 which suggests that the information localization discussed here belongs to the universality class of random percolation on networks \cite{CAH,CAH2,Braun}.

When a fraction $p$ of the nodes is unwilling to reveal information to their neighbors, the critical value $\bar{m}_{c}$ increases strongly  in a non-linear way, as is shown in Fig. \ref{fig3}. As a consequence, when $p$ is large enough it takes unlikely high values of $\bar{m}$ in order to maintain information flow. It is interesting to note that, as in Fig. \ref{fig2}, the result for a random network with $\gamma=2.5$ and $k_{min}=2$ is rather close to that of the internet community network although the last has $k_{min}=1$. 

\begin{figure} 
\begin{center}
\includegraphics[width=8.5cm]{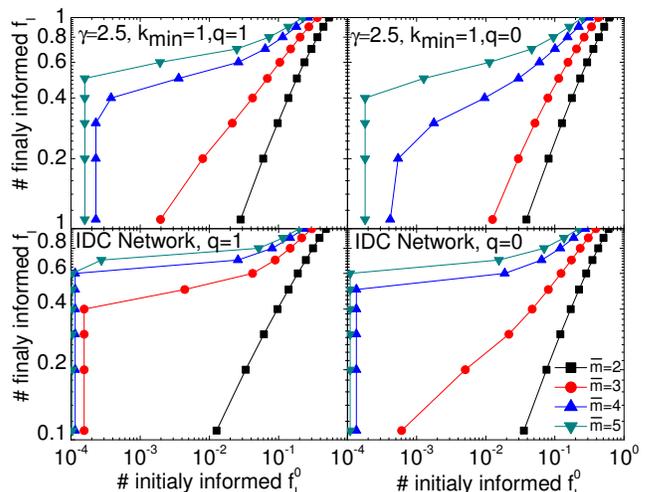}
\end{center}
\vspace{-0.8cm}
\caption{Fraction of eventually informed nodes $f_{I}$ vs the fraction of initially informed nodes $f_{I}^{0}$ for (a) a random  network with $N=10^4, \gamma=2.5, k_{min}=1$,  and $q=1$, (b) the same network as in (a) for $q=0$, (c) The internet dating community (IDC) network ($ N=14782 $)\cite{HEL} and $q=1$ and (d) the same network as in (c) for $q=0$. Curves shown for $\bar{m}$=2 (squares), 3 (circles), 4 (triangles), 5 (reverse triangles).}
\label{fig4}
\end{figure}

Next, we discuss the penetration of relevant (and risky) information into the network when initially not only one node, but a certain fraction $f_{I}^{0}\equiv I_{0}/{N}$ of nodes is informed. 
We are interested in the fraction $ f_{I} \equiv I/N$ of nodes that eventually will be informed. Figure \ref{fig4} shows $ f_{I}$  as a function of $f_{I}^{0}$ for a random scale-free network with $\gamma=2.5$, $k_{min}=1$ and $N=10^4$ nodes and for the IDC network for four values of $\bar{m}$, $\bar{m}=2,3,4,5$. As in the previous figures, we distinguish between the correlated ($q=1$) and the uncorrelated ($q=0$) model.
Below $\bar{m}_{c}$, for example for $\bar{m}=2$, $f_{I}$ increases linearly with $f_{I}^{0}$ except for $f_{I}^{0}$ close to one. The proportionality factor is identical to the mean cluster size $\left\langle s \right\rangle $, since for $ f_{I}^{0}\ll 1$, $I=\left\langle s \right\rangle  I_{0}$. As $\left\langle s \right\rangle $  is small well below $\bar{m}_{c}$, a large fraction of initially informed nodes $f_{I}^{0}$ is needed to inform a large fraction of the network. 
In contrast, for $\bar{m}$ well above $\bar{m}_{c}$, for $\bar{m}=5$ for example, 
$f_{I}$ shows two regimes. A sharp increase at very small values of $f_{I}^{0}$ is followed by a slow increase over nearly the whole $f_{I}^{0}$ regime. The sharp increase is the result of the appearance of a spanning cluster and as a consequence, only a tiny fraction  $f_{I}^{0}$ is needed to inform a considerable number of nodes. The large low-slope region indicates that it is difficult to inform (nearly) all of the network. 
The reason for this is that the fraction of nodes belonging to the infinite cluster is zero at criticality and increases as a power law with increasing distance from the critical point \cite{BH}. Since the majority of nodes is not on the spanning cluster and since at least one initially informed node in every cluster is needed to inform the complete network, informing a large fraction of nodes may become difficult even above $\bar{m}_{c}$. This is evident, for example, in Fig. \ref{fig4} (d) for $\bar{m}=3$ (which is above but close to $\bar{m}_{c}=2.9$). One can see that already for $f_{I}^{0}\simeq 6 \times 10^{-4}$, $f_{I}\simeq 0.1$,i.e. 10\% of the network will be informed but it needs $f_{I}^{0}\simeq 0.1$ in order to inform 50\% of the network. 

Finally, we like to emphasize again that our results are not only valid  for transfer of risky and relevant information but for all those cases where information from one node is likely to be transfered only to a certain number of neighboring nodes. This happens, for example, in product advertisement as (i) someone who is informed of a product will, most likely, talk about it only to those of his acquaintances that he thinks are in need of the product and will appreciate his suggestion (friendly neighbors) and (ii) the receiver of the information has to validate the messanger as a reliable person so that his suggestion is  seriously considered and further promoted. In this case, $f_{I}^{0}$ can be interpreted as a measure of the effort needed to promote a product, which is directly proportional to the promotional cost, while  $\bar{m}$, the average number of people to whom a person will talk of a product, depends on the quality of advertisement since successful advertisement will lead to higher $\bar{m}$ values. Within this interpretation, Fig. \ref{fig4} shows that there is a threshold for the quality of advertisement. Below the threshold, the cost of informing a desired fraction $f_{I}$ of potential consumers is proportional to $f_{I}$ and high. Above the threshold, for a very successful advertisement, a large fraction of potential consumers can be informed at minimum cost.
A further application of our model may be on the spreading of epidemics with very short incubation time. In this case, the epidemics can spread only through the sub-network of those people that are “really” close to the contaminated individual, which is analogous to that of the MTN-network.

{\it Acknowledgment:} We would like to thank Dr. L.K. Gallos and Prof. S. Havlin for stimulating discussions on scale-free networks and Dr. F. Liljeros for providing data on the real social networks used in this study. This work was supported by a European research NEST Project No DYSONET 012911.

\end{document}